\documentstyle[12pt,worldsci]{article}
\begin{document}
\def\z2{\ifmmode Z_2\else $Z_2$\fi}
\def\ie{{\it i.e.},}
\def\eg{{\it e.g.},}
\def\etc{{\it etc}}
\def\etal{{\it et al.}}
\def\ibid{{\it ibid}.}
\def\gev{\,{\rm GeV}}

\def\to{\rightarrow}
\def\epem{\ifmmode e^+e^-\else $e^+e^-$\fi}
\def\Re{{\cal R \mskip-4mu \lower.1ex \hbox{\it e}\,}}
\def\Im{{\cal I \mskip-5mu \lower.1ex \hbox{\it m}\,}}
\pagestyle{empty}
\setlength{\baselineskip}{2.6ex}

\rightline{\vbox{\halign{&#\hfil\cr
&SLAC-PUB-6719\cr
&November 1994\cr
&T/E\cr}}}

\title{{\bf CONSTRAINTS ON THE LEFT-RIGHT SYMMETRIC MODEL FROM $b\to s\gamma$}
\footnote{Work supported by the Department of
Energy, contract DE-AC03-76SF00515.}}
\author{THOMAS G.~RIZZO
\footnote{Presented at the {\it 1994 International Workshop on B
Physics, Physics Beyond the Standard Model at the B Factory},
Nagoya, Japan, October 26-28, 1994.}\\
\vspace{0.3cm}
{\em Stanford Linear Accelerator Center, Stanford University, Stanford, CA
94309, USA}}
\maketitle

\begin{center}
\parbox{13.0cm}
{\begin{center} ABSTRACT \end{center}
{\small\hspace*{0.3cm}

The recent observation by the CLEO Collaboration of the inclusive decay
$b\to s\gamma$ with a branching fraction consistent with the expectations of
the Standard Model is used to constrain the parameter space of the Left-Right
Symmetric Model. Two scenarios are considered: ($i$) equal left- and right-
handed Cabibbo-Kobayashi-Maskawa mixing matrices, $V_L=V_R (~{\rm or}~V_R^*)$
and ($ii$) the Gronau-Wakaizumi model wherein B-decays proceed only via
right-handed currents and $V_L$ and $V_R$ are quite distinct. In the later
case the bounds from $b\to s\gamma$ are combined with other constraints
leaving a parameter range that is very highly restricted and which implies that
this model may soon be completely ruled out by improving data. }}

\end{center}

Rare decay processes allow us to probe energy scales beyond those directly
accessible at current $e^+e^-$ and hadron colliders.
The recent observation of the $b\to s\gamma$ decay by CLEO{\cite {1}}, with a
branching fraction in the range $1-4\cdot10^{-4}$ at $95\%$ CL, coupled to the
possible discovery of the the top quark, with a mass of approximately 175 GeV,
by CDF{\cite {2}}, leads to many restrictions on new physics scenarios beyond
the Standard Model(SM){\cite {3}}. In the analysis below, we consider the
implications of these results for the Left-Right Symmetric Model(LRM)
{\cite {4}}, which is based on the gauge group
$SU(2)_L \times SU(2)_R \times U(1)$.
The `classical' constraints on this model arise from a number of sources
including polarized $\mu$ decay{\cite {5}}, the $K_L-K_S$ mass
difference{\cite {6}}, universality{\cite {7}}, and Tevatron direct $Z',W'$
searches{\cite {8}}. However, the LRM is quite robust and possesses a large
number of free parameters which play an interdependent role in the calculation
of observables and in the constraints resulting from experiment. As far as
$b\to s\gamma$ and the subsequent discussion are concerned there are
essentially 5 parameters of interest:
($i$) $t_\phi=tan \phi$, where $\phi$ is the mixing angle between $W_L$ and
$W_R$ which form
the mass eigenstates $W_{1,2}$, ($ii$) the ratio of masses, $r=M_1^2/M_2^2$,
(with $M_2 \simeq M_R$), ($iii$) the ratio of gauge couplings
$\kappa=g_R/g_L>0.55$, which is expected to be of order unity, ($iv$) the
masses of the right-handed(RH) neutrinos, and ($v$) the elements of the RH
quark mixing matrix, $V_R$. Our discussion below{\cite {9}} assumes
that $W_R$ is the {\it only}
new particle occurring in the $b\to s\gamma$ penguin(\eg  ~charged Higgs
may also contribute to these loops in a highly model dependent way but they
are neglected here) and ignores
any additional phases that may be present in the $W_L-W_R$ mixing matrix that
may arise from complex vev's{\cite {10}}.

We now outline our procedure which to leading order in the QCD corrections
is well known; for details see Ref.9. We first normalize $B(b\to s\gamma)$
to the semileptonic(SL) decay rate using the quark level calculations including
finite phase space and QCD corrections, assuming $m_c/m_b=0.3$ and
$\alpha_s(M_Z)=0.125$. In the LRM, the SL decay rate
is now a function of $t_\phi, r, \kappa$ and the appropriate $V_{L,R}$
factors. Apart from these model parameters, $B$ is expressed in terms of the
coefficients of the $C_{7L}$ and $C_{7R}$ electromagnetic dipole-moment
operators evaluated at the scale
$\mu=m_b$; $m_s$ is assumed to be zero in the results quoted here.
To obtain the numerical values of these operators at the low mass scale
we must know the two $10\times10$ anomalous dimension matrices for the
complete set of operators as well as all the operator coefficients at the weak
scale. To lowest order in $\alpha_s$, only 8 of these coefficients are
non-zero:
\begin{eqnarray}
C_{2L}(M_{W_1}) & = & (1+rt_\phi^2)(V_{cb}V^*_{cs})_L \,, \nonumber \\
C_{2R}(M_{W_1}) & = & \kappa^2(r+t_\phi^2)(V_{cb}V^*_{cs})_R \,, \nonumber \\
C_{10L}(M_{W_1}) & = & \kappa t_\phi(1-r){m_c\over m_b}
(V_{cb}^LV^{*R}_{cs}) \,, \nonumber \\
C_{10R}(M_{W_1}) & = & C_{10L}(M_{W_1}) (L\leftrightarrow R) \,, \\
C_{7L}(M_{W_1}) & = & (V_{tb}V^*_{ts})_L[A_1(x_1)+rt_\phi^2A_1(x_2)]+
{m_t\over m_b}\kappa t_\phi(V_{tb}^RV_{ts}^{*L})[A_2(x_1)-rA_2(x_2)] \,,
\nonumber \\
C_{7R}(M_{W_1}) & = & {m_t\over m_b}\kappa t_\phi(V_{tb}^LV_{ts}^{*R})
[A_2(x_1)-rA_2(x_2)]+\kappa^2(V_{tb}V_{ts}^*)_R[t_\phi^2A_1(x_1)+rA_1(x_2)] \,,
\nonumber
\end{eqnarray}
where $x_{1,2}=m_t^2/M_{W_{1,2}}^2$. The coefficients of the operators
corresponding to the gluon penguin, $C_{8L,R}(M_{W_1})$, can be expressed in
a manner similar to $C_{7L,R}(M_{W_1})$ but with $A_i \to B_i$; note that
both $A_1$ and $B_1$ are the same functions found in the usual SM
calculation. The kinematic functions $A_i$ and $B_i$ are given in Ref.9. An
important feature in
the expressions for $C_{7,8L}$ and $C_{7,8R}$ are terms proportional to
$\kappa t_\phi m_t/m_b$ which arise due to chirality flips and imply that $B$
will be highly sensitive to non-zero values of $t_\phi$ even when $r$ is quite
small. This will be seen explicitly in our results below. The largest new
contribution to $b\to s\gamma$ in the LRM is thus due to the SM $W_L$
picking up a small RH coupling via mixing and vice versa for the $W_R$.

To proceed further we need to make some assumptions about the LRM parameters.
The first case we consider, which one may think is the most natural, is when
$V_L=V_R ~{\rm or} ~V_R^*$ with heavy RH neutrinos and where we know that
$M_R>1.5$ TeV from the $K_L-K_S$ mass difference{\cite {6}}. In Fig.1a
we see the prediction for $B$ as
a function of $t_\phi$ in this case for various values of $m_t$ assuming
$\kappa=1$ and $M_R=1.6$ TeV. To satisfy the CLEO data only a restricted range
of $t_\phi$ is allowed; note the rather weak sensitivity to $m_t$. Fixing
$m_t=175$ GeV and varying $M_R$ we see from Fig.1b that the $t_\phi$
constraints are not sensitive to variations in the $W_R$ mass. If we vary
$\kappa$ for $m_t$ and $M_R$ fixed we obtain Fig.1c which shows that the
$t_\phi$ bounds are quite sensitive to $\kappa$. Note however that if we
consider $B$ as a function of the combination $\kappa t_\phi$ (which enters
directly into the expressions for the weak scale coefficients) there is very
little additional $\kappa$ sensitivity. Thus we see that in this $V_L=V_R$
case the bounds
we obtain on $t_\phi$ are more restrictive than those obtained from either
$\mu$ decay or universality.

In principle, if we give up the assumption that $V_L=V_R(V_R^*)$ there is
very little guidance as to what form $V_R$ might take and bizarre scenarios
may in fact be realized. One possibility is the model of Gronau and
Wakaizumi(GW){\cite {11}} (and several of its clones{\cite {12}}). In this
class of models, $B$ decays proceed {\it only} via RH-currents with the
apparent smallness of $V_{cb}$ explained by the larger $W_R$ mass. The exact
forms taken for $V_{L,R}$ are somewhat model dependent; in the
original GW model, one has
\begin{eqnarray}
V_L & = & \left( \begin{array}{ccc}
1 & \lambda & 0 \\
-\lambda & 1 & 0 \\
0 & 0 & 1
\end{array} \right) \,, \\
V_R & = & \left( \begin{array}{ccc}
c^2 & -cs & s \\
{s(1-c)\over\sqrt 2} & {c^2+s^2\over\sqrt 2} & {c\over\sqrt 2} \\
{-s(1+c)\over\sqrt 2} & -{c-s^2\over\sqrt 2} & {c\over\sqrt 2}
\end{array} \right) \,, \nonumber
\end{eqnarray}
where $\lambda (\simeq 0.22)$ is the Cabibbo angle and $s \simeq 0.09$ and
$c^2=1-s^2$.
Explicitly, to satisfy the $B$ lifetime constraint we must also have
\begin{equation}
M_{W_R}\leq 416.2\, \kappa\left[ {|V^R_{cb}|\over\sqrt 2}\right]^{1/2} \gev
\simeq 415\, \kappa\gev \,,
\end{equation}
which arises from recent determinations of $V_{cb}$ in the SM.
With the assumed forms of $V_{L,R}$ in this model the usual $K_L-K_S$
constraint on the $W_R$ mass is easily circumvented.
In addition, to satisfy the most stringent polarized $\mu$ decay data, the
RH neutrino must be sufficiently massive ($\simeq 17-50$ MeV) but this has
little effect on  B decay itself. (Note that some of the weaker $\mu$ decay
constraints remain.)
Of course, a $W_R$ satisfying the above constraint is relatively
light and should have a significant production cross section at the Tevatron
given the above form of $V_R$. We will assume $M_R=400\kappa$ GeV in what
follows as we will want $M_R$ to be as large as possible. Figs.1d and 1e show
the predicted value of $B$ as a function of
$t_\phi$ for $\kappa=1.5$ and 2, respectively, for different values of $m_t$.
In either case and for all $m_t$ values we see that agreement with the CLEO
result demands that $t_\phi$ lie within either of two very narrow bands with a
magnitude less than 0.001. These general results are maintained at the
semi-quantitative level in the various clones of the GW model{\cite {12}}.
Fixing $m_t=175$ GeV, we see in Fig.1f the overall behaviour of $B$ in the GW
model as $\kappa$ is allowed to vary. As in the previous case, a plot of $B$
as a function of the combination $\kappa t_\phi$ shows little additional
$\kappa$ sensitivity. Thus the CLEO result forces us to
fine-tune $t_\phi$ to a narrow range of very small values in this model.

The GW model uses heavy RH $\nu$'s to avoid the bulk of the $\mu$ constraints.
However, it cannot escape from $\tau$ decay in a similar manner, \ie ~by making
the RH $\nu_\tau$ heavy. Both ALEPH and L3 have measured the branching
fraction for $B\to \tau\nu X${\cite {13}} and found it to be in agreement with
the expectations of the SM{\cite {14}}. If the RH $\nu_\tau$ were heavy
enough to allow the GW
model to escape the $\tau$ decay constraints, this branching fraction would be
seriously compromised as is shown in Fig.2a. We see from the figure that the RH
$\nu_\tau$ must have a mass less than about 0.3 GeV to maintain agreement
with the ALEPH/L3 data implying
that RH currents must be present in $\tau$ decay in the GW model. ALEPH and
L3 have also recently updated the determinations of the Michel parameters
for $\tau$
decay{\cite {15}} which are sensitive to such RH interactions and lead to new
constraints on the GW model(taking $t_\phi \simeq 0$ as we learned from
$b\to s\gamma$). These new constraints, together with those from $\mu$ decay,
direct Tevatron searches, and the $B$ lifetime are combined in Fig.2b. We see
that the GW model parameter space was comfortably large before the recent CDF
$W'$ search and LEP $\tau$ Michel parameter results were
announced{\cite {8,15}}.
The new data highly compresses the model parameter space into the region
near $M_R=800$ GeV with $\kappa=2$. Even this small region will soon become
disallowed if the CDF limit scales logarithmically with increasing integrated
luminosity{\cite {16}}(perhaps in a matter of months). Fig.2b shows the power
of combining rare decay data, precision measurements, and direct searches to
constrain the new physics in the GW version of the LRM.

As in the case of other new physics scenarios, $b\to s\gamma$ has been found
to provide important constraints on the parameters of the LRM.

\vspace{1.0cm}
%
\def\MPL #1 #2 #3 {Mod.~Phys.~Lett.~{\bf#1},\ #2 (#3)}
\def\NPB #1 #2 #3 {Nucl.~Phys.~{\bf#1},\ #2 (#3)}
\def\PLB #1 #2 #3 {Phys.~Lett.~{\bf#1},\ #2 (#3)}
\def\PR #1 #2 #3 {Phys.~Rep.~{\bf#1},\ #2 (#3)}
\def\PRD #1 #2 #3 {Phys.~Rev.~{\bf#1},\ #2 (#3)}
\def\PRL #1 #2 #3 {Phys.~Rev.~Lett.~{\bf#1},\ #2 (#3)}
\def\RMP #1 #2 #3 {Rev.~Mod.~Phys.~{\bf#1},\ #2 (#3)}
\def\ZP #1 #2 #3 {Z.~Phys.~{\bf#1},\ #2 (#3)}
\def\IJMP #1 #2 #3 {Int.~J.~Mod.~Phys.~{\bf#1},\ #2 (#3)}
\bibliographystyle{unsrt}

\begin{thebibliography}{99}
\bibitem{1}
E.H.\ Thorndike, CLEO Collaboration, to appear in the {\it Proceedings of
the 27th International Conference on High Energy Physics}, Glasgow, 1994.
\bibitem{2}
F.\ Abe \etal, CDF Collaboration, \PRL 73 225 1994 ~and \PRD D50 2956 1994 .
\bibitem{3}
J.L.\ Hewett, SLAC-PUB-6521, 1994.
\bibitem{4}
For a review and original references, see R.N.\ Mohapatra, {\it Unification
and Supersymmetry}, (Springer, New York, 1986).
\bibitem{5}
A.\ Jodiddo \etal, \PRD D34 1967 1986  ~and \PRD D37 237 1988 ;
J.\ Imazoto \etal, \PRL 69 877 1992 .
\bibitem{6}
See, for example, G.\ Beall, M.\ Bander, and A.\ Soni, \PRL 48 848 1982 .
\bibitem{7}
A.\ Sirlin, NYU report NYU-TH-93/11/01, 1993.
See also P.\ Langacker and S.U.\ Sankar, \PRD D40 1569 1989 ~for a review of
all constraints.
\bibitem{8}
K.\ Maeshima \etal, CDF Collaboration, Fermilab report FNAL-CONF-94-227-E,
1994.
\bibitem{9}
For details, see T.G.\ Rizzo, \PRD D50 3303 1994 .
\bibitem{10}
Other possibilities have been considered by
K.\ Fujikawa and A.\ Yamada, \PRD D49 5890 1994 ;
P.\ Cho and M.\ Misiak, \PRD D49 5894 1994 ;
K.S.\ Babu, K.\ Fujikawa, and A.\ Yamada, \PLB B333 196 1994 .
\bibitem{11}
M.\ Gronau and S.\ Wakaizumi, \PRL 68 1814 1992 .
\bibitem{12}
W.-S.\ Hou and D.\ Wyler, \PLB B292 364 1992 ;
T.\ Hattori \etal, Tokushima University report 93-06, 1993.
\bibitem{13}
ALEPH Collaboration, presented at the {\it 27th International Conference
on High Energy Physics}, Glasgow, 1994;
M.\ Acciarri \etal, L3 Collaboration, \PLB B332 201 1994 .
\bibitem{14}
A.\ Falk, Z.\ Ligeti, M.\ Neubert and Y.\ Nir, \PLB B326 145 1994 .
\bibitem{15}
For a summary, see J.\ Harton, ALEPH Collaboration, talk presented at the
{\it Workshop on the Tau-Charm Factory in the Era of B-Factories and CESR},
SLAC, August 15-16, 1994.
\bibitem{16}
T.G.\ Rizzo, \PRD D50 325 1994 .

%
\end{thebibliography}

\newpage

{
\noindent
Fig.~1: (a)$B$ as a function of $t_\phi$ assuming $\kappa=1$, $M_R=1.6$ TeV,
and $V_L=V_R$ for $m_t=140(160,180,200)$ GeV as represented by the
dotted(dashed,dash-dotted, solid)curve. (b)Same as (a) but with $m_t=175$ GeV
and $M_R$ varied between 1 and 3 TeV. (c)Same as (a) but with $m_t=175$ GeV and
$\kappa$=0.6(0.8,1,1.2,1.4) corresponding to the dotted(dashed, dash-dotted,
solid, square-dotted)curve. (d)Same as (a) but in the GW model with $M_R=600$
GeV and $\kappa=1.5$. (e)Same as (d) but with $M_R=800$ GeV and $\kappa=2$.
(f)Same as (d) but with $m_t=175$ GeV and $M_R=400\kappa$ GeV with $\kappa$
varying between 1(outer curve) and 2(inner curve) in steps of 0.2.

\medskip

\noindent
Fig.~2: (a)Branching fraction for the decay $B\to \tau\nu_R X$ in the GW model
as a function of the mass of the RH-neutrino. The combined ALEPH+L3 $95\%$ CL
lower bound is the horizontal dashed line. (b)Constraints on $\kappa$ and
$M_R$ in the GW model: the solid line is the upper bound from Eq.3. The
dash-dotted line is the lower bound from $\tau$ and $\mu$ decay data. The
dotted(dashed) line is the CDF lower bound from the '88-'89 run (run 1a). The
currently allowed region lies in the upper right hand corner.
}

\end{document}